\documentclass[%
 aip,
 amsmath,amssymb,
 reprint,%
]{revtex4-1}

\usepackage{graphicx}
\usepackage{dcolumn}
\usepackage{bm}

\usepackage[hidelinks]{hyperref}
\usepackage{comment}
\usepackage{braket}
\usepackage{xcolor}
\usepackage{colortbl}
\usepackage{listings}

\usepackage[utf8]{inputenc}
\usepackage[T1]{fontenc}
\usepackage{mathptmx}
\usepackage{etoolbox}

\makeatletter
\def\@email#1#2{%
 \endgroup
 \patchcmd{\titleblock@produce}
  {\frontmatter@RRAPformat}
  {\frontmatter@RRAPformat{\produce@RRAP{*#1\href{mailto:#2}{#2}}}\frontmatter@RRAPformat}
  {}{}
}%
\makeatother

\begin{document}

\preprint{AIP/123-QED}

\title[HORTENSIA]{HORTENSIA, a program package for the simulation of nonadiabatic autoionization dynamics in molecules}

\author{Kevin Issler}
\affiliation{Julius-Maximilians-Universität Würzburg, Institut f\"{u}r Physikalische und Theoretische Chemie, Emil-Fischer-Str. 42, 97074 Würzburg, Germany}
 
\author{Roland Mitrić}
 \email{roland.mitric@uni-wuerzburg.de}
\affiliation{Julius-Maximilians-Universität Würzburg, Institut f\"{u}r Physikalische und Theoretische Chemie, Emil-Fischer-Str. 42, 97074 Würzburg, Germany}

\author{Jens Petersen}
\email{jens.petersen@uni-wuerzburg.de}
\affiliation{Julius-Maximilians-Universität Würzburg, Institut f\"{u}r Physikalische und Theoretische Chemie, Emil-Fischer-Str. 42, 97074 Würzburg, Germany}

\date{\today}

\begin{abstract}
We present a program package for the simulation of ultrafast vibration-induced autoionization dynamics in molecular anions in the manifold of the adiabatic anionic states and the discretized ionization continuum. This program, called HORTENSIA ($\underline{Ho}$pping $\underline{r}$eal-time $\underline{t}$rajectories for $\underline{e}$lectron-ejection by $\underline{n}$onadiabatic $\underline{s}$elf-$\underline{i}$onization in $\underline{a}$nions), is based on the nonadiabatic surface-hopping methodology, wherein nuclei are propagated as an ensemble along classical trajectories in the quantum-mechanical potential created by the electronic density of the molecular system. The electronic Schrödinger equation is numerically integrated along the trajectory, providing the time evolution of electronic state coefficients, from which switching probabilities into discrete electronic states are determined. In the case of a discretized continuum state, this hopping event is interpreted as the ejection on an electron. The derived diabatic and nonadiabatic couplings in the time-dependent electronic Schrödinger equation are calculated from anionic and neutral wavefunctions obtained from quantum chemical calculations with commercially available program packages interfaced with our program.

Based on this methodology, we demonstrate the simulation of autoionization electron kinetic energy spectra that are both time- and angle-resolved. In addition, the program yields data that can be interpreted easily with respect to geometric characteristics such as bonding distances and angles, which facilitates the detection of molecular configurations important for the autoionization process.

Moreover, useful extensions are included, namely generation tools for initial conditions and input files as well as for the evaluation of output files both through console commands and a graphical user interface.
\end{abstract}

\maketitle

\section{Introduction}

After generation of a temporary molecular anion through electron attachment, there are three possible competing relaxation mechanisms.\cite{illenbergerBook} 
These are a) radiative deactivation, assuming that there is a lower-lying anion state that is stable with respect to ionization, b) dissociative electron attachment, in which the captured electron induces geometric change in the molecule resulting in fragmentation into more stable products, a neutral and an anionic subsystem.
And lastly, c) autoionization, in which after a finite period of time the metastable state decays via electron ejection.
The process of dissociative electron attachment is observed for example in DNA, where capture of low-energy electrons leads to single and double strand breaks\cite{dnascience, dnaprl}, or in a variety of substances in nanoscale thin films\cite{arumainayagam2010}. 
Prominent examples for autoionization include excited dipole- and quadrupole-bound anions with binding energies slightly below the ionization threshold\cite{wang2013, wang2014, wang2017, wang2022}, intermolecular Coulombic decay at the FADH$^-$ cofactor involved in DNA-photolesion repair\cite{harbach2013} and autoionization induced by vibrational excitation in organic molecules\cite{adams2010, bull2016, devine2017, devine2018, adams2019, anstoeter2020}. 
Generally the finite lifetime of a metastable state with respect to autoionization can vary strongly from only a few femtoseconds\cite{naff1968, jordan1987} up to milliseconds\cite{naff1968, suess2002}. 
Recently, several experiments have provided insights into the dynamics of such processes in dipole- and quadrupole-bound organic anions on a (sub-)picosecond timescale.\cite{bull2016, bull2019, anstoeter2020, verlet2020, kang2020, kang2021}

Although the process of autoionization is well-known and -observed experimentally by a multitude of methods, as can be seen in the references given above, the theoretical description of autoionizing systems is challenging\cite{herbert2015}, especially if one is interested in the mechanistic details of the intricate ultrafast relaxation dynamics. 
Autoionization processes can follow different general mechanisms, depending on how energy is redistributed among the system's degrees of freedom. Besides a purely electronic variant, where already the electronic energy of the system lies above the ionization threshold and electron ejection may proceed via tunneling, there is also the possibility of a nonadiabatic mechanism in which rotational or vibrational energy of the nuclei is transformed into the kinetic energy of the ejected electron. 

In the following, we focus on the case of vibrational autoionization. This process can thus be viewed as a nonadiabatic transition between a vibrationally excited bound N-electron system and continuum electronic states consisting of an N-1 electron molecular core and a free electron. Early theoretical treatments have focused on the computation of ionization rates\cite{berry66,simons1984,simons99} as well as on establishing propensity rules for the ionization transitions\cite{simons1981}. While a full dynamical treatment of vibrational autoionization is highly desirable, an entirely quantum-dynamical approach is computationally prohibitive. As an alternative, a mixed quantum-classical ansatz can be considered, further motivated by the success of this type of methodology in the description of bound-state nonadiabatic processes and the simulation of time-resolved spectroscopic signals.\cite{humeniuk2013, persico2014, worth2015, barbatti2018, roeder2019} Although to date there have been several implementations of mixed quantum-classical dynamics simulations for bound-state problems made publicly available\cite{newtonx,sharc,jade}, no program addressing the simulation of vibration-induced autoionization processes has been published so far.

Therefore, in this work we present the program package implementing our approach to describe vibrational autoionization through quantum-classical dynamics in the framework of the surface-hopping methodology in the manifold of bound and continuum electronic states as described recently\cite{aid}.
Therein, nuclear motion is considered classically, while the electronic system is treated quantum-mechanically. 
Nonadiabatic transitions between electronic states accompanied by change of the classical vibrational energy of the molecule describe the energy exchange between the two subsystems. 
With this program package and the underlying methodology, one is able to gain insight into the geometric and electronic evolution in the course of the autoionization process as well as to calculate the time-, energy- and angle-distribution of the generated free electrons, which serve as experimental observables for monitoring autoionization dynamics.

We illustrate our program on the example of the 2-cyanopyrrolide anion, which bears a dipole-bound excited state slightly below the electron detachment threshold while the vibrationally excited states are metastable and decay via autoionization.\cite{wang2022}

In the following section a brief theoretical description of the method is given. In section \ref{section:program} an overview of the actual implementation is provided. The subsequent section \ref{section:discussion} details performance-related issues, namely quality of approximations in the theory and runtime and memory optimization within the program, as well as a dynamics simulation example for the 2-cyanopyrrolide anion. Finally in section \ref{section:conclusion} a conclusion and outlook are given.

\section{Theory}
\label{section:theory}

Our methodological framework is based on the surface-hopping procedure as proposed by Tully\cite{tully1990}, in which the coupled electron-nuclear dynamics of molecular systems is approached in a quantum-classical fashion. 
Specifically, the nuclei are propagated classically according to Newton's equations of motion,
\begin{align}
M\ddot{\textbf{R}}(t)
=
\mathbf{F}_i(\mathbf{R}[t])
\equiv
-\nabla_{\textbf{R}} E_i(\textbf{R}[t]),\label{newton}
\end{align}
where the force $\mathbf{F}_i(\mathbf{R}[t])$ is obtained as the negative gradient of the electronic potential energy surface (PES) $E_i(\textbf{R}[t])$. In the above equation, $M$ denotes a diagonal matrix containing the nuclear masses. 
For an ensemble of initial conditions, this leads to trajectories $\textbf{R}(t)$ moving on the given PES.
Simultaneously, the electronic time-dependent Schrödinger equation
\begin{align}
i\hbar \dot {\Psi}(\textbf{r};\textbf{R}[t])
=
\hat{H}_{el} \Psi(\textbf{r};\textbf{R}[t])
,\label{el_schroedinger}
\end{align}
with the electronic Hamiltonian $\hat{H}_{el}$ is solved. 
The electronic wavefunction can be expanded into a set of orthonormal basis states, which in the case of autoionization includes bound states $\Phi_{m'}$ (denoted with a primed index) as well as continuum states $\tilde{\Phi}_{m''}$ (denoted with a double-primed index):
\begin{align}
\Psi\big(\textbf{r},\textbf{R}[t],t\big) 
=& 
\sum_{m'} c_{m'}(t) \Phi_{m'}\big(\textbf{r},\textbf{R}[t]\big)\ 
+\nonumber\\
&\sum_{m''} \int\! d^3\textbf{k}\ \tilde{c}_{m''}(\textbf{k},t) \tilde{\Phi}_{m''}(\textbf{k},\textbf{r},\textbf{R}[t]),
\label{expansionWF}
\end{align}
where \textbf{k} denotes the continuously varying wave vector of the free electron, while $m''$ is the quantum number of the remaining neutral state.
We assume the wavefunctions $\Phi_{m'}$ and $\tilde{\Phi}_{m''}$ to be single Slater determinants (ground state) or an expansion of singly excited Slater determinants (excited state).
In the frame of the presented methodology we discretize the continuum states, leading to
\begin{align}
&\int d^3\textbf{k}\,
\tilde{c}_{m''}(\textbf{k},t)
\tilde{\Phi}_{m''}(\textbf{k},\textbf{r},\textbf{R}[t])\nonumber\\
&\approx 
\sum_i 
    (\Delta {\cal V}_k)^\frac{1}{2} \tilde{c}_{m''}(\textbf{k}_i,t) 
    (\Delta {\cal V}_k)^\frac{1}{2} \tilde{\Phi}_{m''}(\textbf{k}_i,\textbf{r},\textbf{R}[t])\nonumber\\
&\approx  
\sum_i 
    c_{m''}(\textbf{k}_i,t)  
    \Phi_{m''}(\textbf{k}_i,\textbf{r},\textbf{R}[t]),
\end{align}
where $\Delta {\cal V}_k$ denotes the volume element in $k$-space and the discretized and continuum state expansion coefficients are related according to $c_{m''}(\textbf{k}_i,t)=(\Delta {\cal V}_k)^\frac{1}{2} \tilde{c}_{m''}(\textbf{k}_i,t)$. The actual determination of the wave vectors and the implementation of the discretization procedure are explained in detail in the next chapter.

Insertion of Eq. (\ref{expansionWF}) into the time-dependent Schrödinger equation (\ref{el_schroedinger}), multiplication from the left by an eigenstate $\bra{\Phi_n}$ and evaluation of the arising terms leads to a set of coupled differential equations for the electronic state coefficients $c_n$:
\begin{align}
\dot{c}_n(t)
=
\sum_j 
    \left[ 
        -\frac{i}{\hbar} H_{nm}(\textbf{R}[t]) - D_{nm} (\textbf{R}[t])
    \right] 
    c_m(t),
\label{schroedinger}
\end{align}
with the matrix elements of the electronic Hamiltonian $H_{nm} = \braket{\Phi_n | H_{el} | \Phi_m}$ and the nonadiabatic couplings $D_{nm} = \braket{\Phi_n | \dot{\Phi}_m} = \dot{\textbf{R}}\cdot \braket{\Phi_n | \nabla_R | \Phi_m}$.
These can be divided into separate expressions for the bound and continuum states, resulting in the diabatic and nonadiabatic couplings between two bound anion states,
\begin{align}
H_{n'm'}
&=
\braket{\Phi_{n'} | \hat{H} | \Phi_{m'}}
\label{hij}\\
D_{n'm'}
&=
\braket{\Phi_{n'} | \dot{\Phi}_{m'}}
,\label{dij}
\end{align}
and between a bound and a discretized continuum state,
\begin{align}
H_{n''m'}(\textbf{k}_i)
&=
(\Delta {\cal V}_k)^\frac{1}{2}
\braket{\tilde{\Phi}_{n''}(\textbf{k}_i) | \hat{H} | \Phi_{m'}}
\label{hik}\\
D_{n''m'}(\textbf{k}_i)
&=
\braket{\Phi_{n''}(\textbf{k}_i) | \dot{\Phi}_{m'}}
=
(\Delta {\cal V}_k)^\frac{1}{2}
\braket{\tilde{\Phi}_{n''}(\textbf{k}_i) | \dot{\Phi}_{m'}}.
\label{dik}
\end{align}
In the above equations, the approximation to neglect the coupling terms between the continuum states has been introduced.
The discretized continuum states consist of an antisymmetrized product of a bound $N-1$ electron neutral state and a molecular scattering state of the free electron
\begin{align}
\tilde{\Phi}_{n''}(\textbf{k}_i)
=
{\cal A} 
\left(
    \Phi^{\textrm{(n)}}_{n''}
    \cdot
    \psi(\textbf{k}_i)
\right).\label{antisymm}
\end{align}
The simplest approximation to the free electron states in the presence of a neutral molecular core are plane waves
\begin{align}
\psi(\textbf{k}_i)\approx {\cal N} \textrm{e}^{i\textbf{k}_i\cdot \textbf{r}}
\end{align}
with a normalization constant ${\cal N} = (2\pi)^{-3/2}$ to satisfy the orthonormality demanded in Eq. (\ref{expansionWF}). 
Since this function would be completely independent on the electronic and nuclear configuration of the molecular core, which is a strong simplification, the plane waves are orthogonalized with respect to the anion's molecular orbitals (MOs) $\phi_m$ to include (at least to a certain degree) dependence on the molecular structure according to
\begin{align}
\tilde{\psi}(\textbf{k}_i)
&= 
(2\pi)^{-3/2}
{\cal N}_{ortho}
\left( 
    \textrm{e}^{i\textbf{k}_i\cdot \textbf{r}} 
    - 
    \sum_m^{\mathrm{occ}} 
        \braket{\phi_m | \textrm{e}^{i\textbf{k}_i\cdot \textbf{r}}} 
        \phi_m
\right )\nonumber\\
&=
{\cal N}_{ortho}
\left(
\psi(\textbf{k}_i)
-
\sum_m^{\mathrm{occ}} 
    \braket{\phi_m | \psi(\textbf{k}_i)}\, 
    \phi_m
\right),
\label{orth_pw}
\end{align}
with the normalization constant
\begin{align}
{\cal N}_{ortho}
=
\left(
    1 
    - 
    \sum_m^{\mathrm{occ}}
        \big|
            \braket{\phi_m | \psi(\textbf{k}_i)}
        \big|^2
\right)^{-1/2}
\label{orthoNorm}
\end{align}
arising from the orthogonalization.

Notably, the summation over $m$ includes the occupied MOs in all 'relevant' Slater determinants of all considered electronic states, that is, we considered all determinants which are needed to sufficiently represent the ground state and full CIS wavefunction of the excited state.
Beginning from the highest contribution to a wavefunction, determinants are included until a specific percentage or a user-adjusted maximum number of configurations per electronic state is reached (95 \% / 10 configurations in the case of vinylidene\cite{aid}).
Considering for now the special case where only the anion's ground state is included, the used MOs are simply the energetically lowest ones up to the highest-occupied molecular orbital (HOMO).

The overlap integral between a plane wave and an MO present in Eq. (\ref{orthoNorm}), $\braket{\phi_m | \psi(\textbf{k}_i)}$, can be computed analytically by expanding the MO into the Gaussian atomic orbital (AO) basis, with the integral involving a single AO $|\nu\rangle$ given by
\begin{align}
\braket{\nu | \psi(\textbf{k})}
=&\ 
(2\pi)^{-3/2}
\int d^3\mathbf{r}\,
\textrm{e}^{i\textbf{k}\cdot \textbf{r}}
\varphi_\nu(\textbf{r})\nonumber\\
=&\ 
(2\alpha_\nu)^{-3/2}
\exp{\left(i\textbf{k}\cdot\textbf{A}_\nu -\frac{k^2}{4\alpha_\nu}\right)}
\nonumber\\
&\times 
\prod_{j=x,y,z}
    (-i\sqrt{4\alpha_\nu})^{-n_{\nu,j}}
    H_{n_{\nu,j}}\left(
        \frac{k_j}{\sqrt{4\alpha_\nu }}
    \right)
\label{ft_gauss},
\end{align}
where the $H_{n_{\nu,j}}$ are the Hermite polynomials of order $n_{\nu,j}$.

\subsection{Electronic coupling terms}

There are anionic systems, for example the vinylidene anion\cite{aid}, that do not support a bound excited state, in which case the consideration of only the ground state and the continuum in the process of autoionization is sufficient. 
Besides that, for example in molecules exhibiting dipole-bound excited states \cite{jordan2003, wang2022, issler2023_3}, several bound anionic states and the interaction among them are relevant as well. 
Nonetheless, to keep the formalism concise, if not noted otherwise we discuss in the following the electronic coupling terms for the special case of both anion and neutral molecule being in their respective electronic ground states, which in turn are represented by a single Slater determinant. 
The generalization to excited states and/or multideterminantal wavefunctions is straightforward.\cite{issler2023_3}
We denote the bound anionic ground state wavefunction by $\ket{\Phi_0}$ and the continuum wavefunctions by $\ket{\Phi_i}$, the latter being constructed as an antisymmetrized product of the neutral ground state and a free electron state function with wave vector $\textbf{k}_i$, similar to Eq. (\ref{antisymm}).

\subsubsection{Diabatic couplings}
In the case of two adiabatic bound anion states, the coupling matrix elements $H_{n'm'}$ given in Eq. (\ref{hij}) yield zero for all $n' \neq m'$ since these states are orthonormal eigenstates of the electronic Hamiltonian.

On the other hand, since in our methodology the bound and continuum state wavefunctions are constructed using separate quantum-chemical calculations for the anion and neutral, and the free electron wavefunction is taken as a plane wave, the continuum state functions are crude approximations to the actual adiabatic eigenfunctions of the electronic Hamiltonian for the $N$-electron system and therefore, diabatic couplings between the bound and continuum electronic states arise.

As elaborated in detail in Ref. \onlinecite{aid}, according to Eq. (\ref{hik}) and defining $V_{i0}^{\mathrm{dia}}(\textbf{k}_i)$ as
\begin{align}
H_{i0}(\textbf{k}_i)
\equiv
\braket{\Phi_i | \hat{H} | \Phi_0}
\equiv
(\Delta {\cal V}_k)^\frac{1}{2}\, 
V^{\mathrm{dia}}_{i0}(\textbf{k}_i),
\end{align}
the diabatic coupling between a bound and a continuum state can be written in terms of the AO basis as
\begin{alignat}{3}
V^{\mathrm{dia}}_{i0}(\textbf{k}_i)
&=&
\sum_{\lambda\mu\nu}
    \Bigg[&
        A_{\lambda\mu\nu}
        \Big(
            \braket{\mathbf{k}_i \lambda || \mu \nu}
            -\sum_{\sigma}
                B_{\sigma}
                \braket{\sigma \lambda || \mu \nu}
        \Big) + \nonumber\\
        &&&
        \bar{A}_{\lambda\mu\nu}
        \Big(
            \braket{\mathbf{k}_i \lambda | \mu \nu}
            -
            \sum_{\sigma}
                B_{\sigma}
                \braket{\sigma \lambda | \mu \nu}
        \Big)
    \Bigg].
\label{vdia_ao_main}
\end{alignat}
In this formula the Greek letters denote the AO basis functions, $\braket{\mathbf{k}_i \lambda | \mu \nu}$ is an electron-electron repulsion integral and $\braket{\mathbf{k}_i \lambda || \mu \nu} = \braket{\textbf{k}_i \lambda | \mu \nu} - \braket{\textbf{k}_i \lambda | \nu \mu}$ its antisymmetrized variant. 
The prefactors $A_{\lambda\mu\nu}$, $\bar{A}_{\lambda\mu\nu}$ and $B_{\sigma}$ comprise AO expansion coefficients and overlap integrals and are defined as follows (assuming that the extra electron of the anion has $\alpha$ spin):
\begin{alignat}{1}
A_{\lambda\mu\nu}
=&
\sum_n^{\mathrm{occ},\alpha} 
    \sum_{q,p<q}^{\mathrm{occ},\alpha} 
        (-1)^{n+p+q-1} 
        \mathrm{det}\ 
        \mathbf{S}_{in,pq}
        \nonumber\\
        &
        \times\left(
            c_\lambda^{(n)} 
            - 
            \sum_u^{\mathrm{occ},\alpha}
                c_\lambda^{(u)} 
                S_{nu}
        \right)
        c_\mu^{(p)} c_\nu^{(q)}
\label{Afactor}
\\
\bar{A}_{\lambda\mu\nu}
=&
\sum_{\bar{n}}^{\mathrm{occ},\beta} 
    \sum_{p}^{\mathrm{occ},\alpha} 
        \sum_{\bar{q}}^{\mathrm{occ},\beta}
            (-1)^{\bar{n}+p+\bar{q}-1} 
            \mathrm{det}\ 
            \mathbf{S}_{i\bar{n},p\bar{q}}
            \nonumber\\
            &
            \times\left(
                c_\lambda^{(\bar{n})} 
                - 
                \sum_{\bar{u}}^{\mathrm{occ},\beta}
                    c_\lambda^{(\bar{u})} 
                    S_{\bar{n} \bar{u}}
            \right)
            c_\mu^{(p)} c_\nu^{(\bar{q})}
\label{Abarfactor}
\\
B_{\sigma}
=&
\sum_r^{\mathrm{occ},\alpha}
    \sum_\rho
        c_\sigma^{(r)} 
        c_\rho^{(r)}
        \braket{\textbf{k}_i | \rho},
\label{Bfactor}
\end{alignat}
where the indices (including their variants with an overbar) $p,q,r$ refer to anion MOs, $n,u$ to neutral MOs, and $\mathrm{det}\ \mathbf{S}_{in,pq}$ denotes the minor determinant of the overlap matrix between continuum and bound state orbitals where the rows of the free electron orbital $\tilde{\psi}(\mathbf{k}_i)$ and neutral orbital $\chi_n$ as well as the columns of anion orbitals $\phi_p$ and $\phi_q$ have been deleted.
For the full derivation of these equations the reader is referred to Ref. \onlinecite{aid}.

\subsubsection{Nonadiabatic couplings}
The nonadiabatic coupling terms as defined in Eqs. (\ref{dij}) and (\ref{dik}) are calculated using the finite-difference approximation for the time derivative, which leads to
\begin{align}
D_{i0}(t)
&=
\braket{\Phi_i(t) | \frac{d}{dt}{\Phi}_0(t)}\\  
&\approx
\frac{1}{2\Delta t} 
\Big( 
    \braket{\Phi_i(t-\Delta t) | \Phi_0(t)} 
    - 
    \braket{\Phi_i(t) | \Phi_0(t-\Delta t)} 
\Big)\label{nonad}
\end{align}
In the case of two anionic bound states, these terms are evaluated according to Refs. \onlinecite{mitric2008, werner2008, werner2010}.

One can simplify the arising terms by integrating over all but one electron coordinate. For the first term of Eq. (\ref{nonad}) this yields
\begin{align}
\braket{\Phi_i(t') | \Phi_0(t)}
=
N^{-1/2}
\braket{\tilde{\psi}(\textbf{k}_i,t') | \psi^D(t',t)}
,\label{pw_dyson}
\end{align}
where we have abbreviated $t'=t-\Delta t$ and have defined the one-electron function $\psi^D(t',t)$, which is an analog to a molecular Dyson orbital with the $N$- and $N-1$- wavefunctions taken at different time steps and geometries.
Using Eqs. (\ref{orth_pw}) and (\ref{pw_dyson}) the resulting nonadiabatic coupling terms read
\begin{widetext}
\begin{align}
D_{i0}(\textbf{k}_i,t) 
= 
\frac{(\Delta {\cal V}_k)^\frac{1}{2}{\cal N}_{ortho}}{2 \sqrt{N} \Delta t} 
\Big[ 
    \braket{\psi(\textbf{k}_i) | \psi^D(t',t)}
    -
    \braket{\psi(\textbf{k}_i) | \psi^D(t,t')}
    &-
    \sum_n 
        \braket{\psi(\textbf{k}_i) | \phi_n(t)}\,
        \braket{\phi_n(t') | \psi^D(t',t)} 
    \nonumber\\
    &+
    \sum_n 
        \braket{\psi(\textbf{k}_i) | \phi_n(t)}\,
        \braket{\phi_n(t) | \psi^D(t,t')} 
\Big]. 
\label{nonad_pw_mo}
\end{align}
\end{widetext}

\subsection{Adiabatic ionization and electronic decay}
\label{subsection:el_ion}

The main focus of the above presented methodology lies on describing the nonadiabatic process of vibrational autoionization. However, in the course of the molecule's dynamical evolution instances can occur where the occupied anionic state becomes unbound as the result of changes in nuclear geometry.
In this case, ionization is possible as an exclusively \textit{adiabatic} electronic process without coupling to the nuclear motion.
This process can be included approximately in our method by simulating the temporal spread of the ejected electron as a wavepacket evolving freely in space. As a quantitative measure, the electronic spatial extent, i.e., the expectation value of $\hat{\mathbf{r}}^2$, is calculated as a function of time.

Specifically, once a time step is reached where the VDE has become negative, the highest-occupied orbital of the last bound geometry, $\phi(\textbf{r}, t_0)$, is used as the initial free electronic wavepacket. 
In the case where one only considers the anionic ground state, this corresponds to the HOMO. 
If also an excited state is involved, natural transition orbitals (NTOs)\cite{martin2003} are calculated and the highest-occupied and lowest-unoccupied NTO (HONTO and LUNTO) are used for the anionic ground and excited state, respectively.
Such an electronic wavepacket is then propagated in time and its spatial extent is evaluated according to
\begin{align}
\braket{\hat{\mathbf{r}}^2}(t) 
&= 
\braket{\phi(\mathbf{r},t) |\hat{\mathbf{r}}^2 |\phi(\mathbf{r},t)} 
\nonumber\\
&=
\sum_{\mu\nu} c_\mu c_\nu 
\braket{\varphi_\mu(\mathbf{r},t) | \hat{\mathbf{r}}^2 | \varphi_\nu (\mathbf{r},t)}.
\label{expect_r2}
\end{align}
Here $\varphi_{\mu, \nu}$ denote the Gaussian atomic basis functions freely propagated in time:
\begin{equation}
\varphi_\mu(\mathbf{r},t) = \int d^3\mathbf{r}'\, K(\mathbf{r},\mathbf{r}',t,0) \varphi_\mu (\mathbf{r}',0)
\end{equation}
with the free electron propagator
\begin{equation}
K(\mathbf{r},\mathbf{r}',t,0)
=
\Braket{\mathbf{r} | \mathrm{e}^{-i\hat{\mathbf{p}}^2 t/2m_e\hbar}|\mathbf{r}'}.
\end{equation}
Using Cartesian Gaussian basis functions of $s$, $p$ and $d$ type one obtains the following analytic expression for the electronic wavepacket:
\begin{align}
\varphi_\mu(\mathbf{r},t) 
= 
N_{l_xl_yl_z} \mathrm{e}^{-\frac{\alpha}{1+i\beta t}\mathrm{r}^2}
\left[
-\Lambda \frac{i\beta t}{2\alpha}
(1+i\beta t)^{-\frac{5}{2}}
+\right.\nonumber\\
\left.
(1+i\beta t)^{-\frac{3}{2} - \sum_j l_j} 
\prod_{j=x,y,z} (r_j - A_j)^{l_j}\right],
\end{align}
where $\textbf{A}$ is the spatial center of the respective basis function, $l_i$ denotes the angular momentum quantum number for the $i$'th spatial direction and $\Lambda$ is a constant that is unity if one of the $l_i = 2$ and zero if all $l_i<2$.
The AO integrals in Eq. (\ref{expect_r2}) are calculated with an implementation of the McMurchie-Davidson scheme\cite{mcmurchie1978}.
To relate the spatial extent in a simple way to the lifetime of the unbound state, an auxiliary spherically symmetric electron distribution is considered which within the initially determined radius $r_0=\sqrt{\braket{\textbf{r}^2}(t_0)}$ contains a probability of 99\%. Subsequently, with $\braket{\textbf{r}^2}$ increasing with time, the probability within $r_0$ decreases, giving rise to a population decay curve which can be related to a time constant $\tau$.
The latter is incorporated into the propagation of the electronic wavefunction given by Eq. (\ref{schroedinger}) by adding an imaginary component to the electronic state energy,

\begin{equation}
E^\mathrm{(a)}
\rightarrow 
E^\mathrm{(a)}-\frac{i\hbar}{2\tau},
\end{equation}
which leads to an exponential population decay due to adiabatic ionization in regions where the VDE is negative for the given electronic state.

\subsection{Surface-hopping procedure}
Solution of the set of Eqs. (\ref{schroedinger}) along a nuclear trajectory yields the time-dependent electronic state coefficients $c_n(t)$. 
Within the surface-hopping methodology, a switch from the occupied bound electronic state $n$ to any other state $m$ is determined by the hopping probability depending on the electronic state populations $\rho_{nn} = |c_n|^2$, which is 
\begin{align}
P_{n\rightarrow m}
= 
-\frac{\dot{\rho}_{nn}}{\rho_{nn}} 
\frac{\dot{\rho}_{mm}}{\sum_k \dot{\rho}_{kk}} \Delta t
\label{probs}
\end{align}
for $\dot{\rho}_{nn} < 0$ and $\dot{\rho}_{mm} > 0$ and zero in any other instance. In the above expression, the sum over $k$ includes all states with $\dot{\rho}_{kk}>0$. 
In case a surface hop occurs, to ensure energy conservation the nuclear velocities are rescaled such that for kinetic energies $T$ and electronic potential energies $E_n$ of anion (a) and neutral (n) the following conditions are fulfilled:
\begin{align}
T'^{\textrm{(a)}}
=
T^{\textrm{(a)}} +
E_n^{\textrm{(a)}} - 
E_m^{\textrm{(a)}}
\end{align}
for a hop between anionic bound states and
\begin{align}
T'^{\textrm{(n)}}
=
E_n^{\textrm{(a)}} + 
T^{\textrm{(a)}} - 
E_m^{\textrm{(n)}} - 
E_{\textrm{el}}(\textbf{k}_i)
\end{align}
for a hop into the continuum (i.e. autoionization).
For a more detailed description of the hopping procedure the reader is referred to Ref. \onlinecite{domckebook}.

\section{Program implementation}
\label{section:program}

\begin{figure*}[htb]
 \centering
 \includegraphics[width=\textwidth, draft=false]{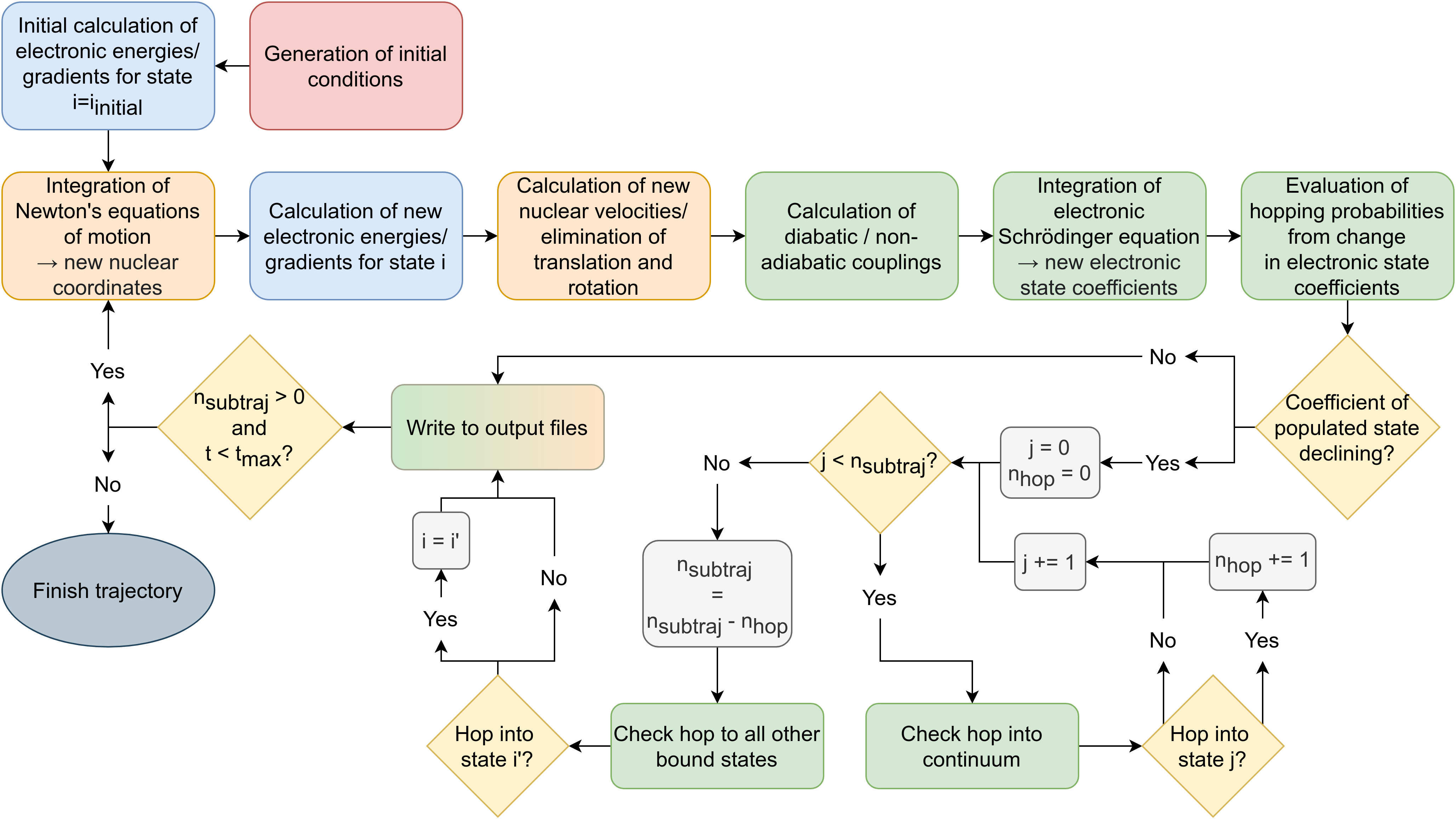}
 \caption{Schematic of the dynamics procedure as implemented in the HORTENSIA program package. The box coloration matches the specific tasks to the program modules as follows: red: wigner/wignerEnsemble.py, blue: external quantum-chemistry program, orange: nuclearDynamics.py, green: populationDynamics.py}
 \label{fgr:flowchart}
\end{figure*}

In the following chapter a detailed account of how the theory is actually implemented in the program package will be provided.
For an easier understanding, in Fig. \ref{fgr:flowchart} the program flow is displayed schematically, with a color code indicating the module handling the respective task.

Starting from the generation of an ensemble of nuclear coordinates $\textbf{R}(t)$ and velocities $\dot{\textbf{R}}(t)$ at the time $t = t_{initial}$ using the \texttt{wignerEnsemble} module in the \texttt{wigner} folder (red), a first quantum-chemical calculation is performed by an external quantum-chemistry program - to date these include Gaussian09/Gaussian16 \cite{g16} and QChem \cite{qchem} (blue) - which yields the forces from which the accelerations $\ddot{\textbf{R}}(t)$ of the nuclei are computed. 
The nuclei are then propagated by integration of Newton's equations of motion for one nuclear time step using the \texttt{nuclearDynamics} module (orange).
With the new nuclear coordinates $\textbf{R}(t + \Delta t)$, a new set of quantum-chemical calculations can be performed, yielding the new energy gradients necessary for the evaluation of the velocities $\dot{\textbf{R}}(t + \Delta t)$. 
With the quantum-chemical calculations at $t$ and $t + \Delta t$, one is now able to construct the electronic continuum states as well as the coupling matrices of the diabatic and nonadiabatic couplings using the \texttt{populationDynamics} module (green). 
From this point, the electronic state coefficients $\textbf{c}(t)$ are propagated in parallel to the nuclear dynamics by integrating the electronic Schrödinger equation, yielding $\textbf{c}(t + \Delta t)$. 
These are utilized to compute hopping probabilities from the occupied bound state to all other (bound and continuum) states. 
The switching between the states is induced stochastically according to the respective hopping probabilities given in Eq. (\ref{probs}). 
After writing the results into the various output files time is shifted to $t = t + \Delta t$, thereby completing one time step. 

To make this initial overview more specific, in the following the underlying algorithms are explained in more detail.

\subsection{Electronic structure calculation}

All electronic structure and energy gradient calculations can  be performed by using any Kohn-Sham (TD)-DFT level of theory provided within the Gaussian09, Gaussian16 or QChem program packages. 
The AO basis set needs to be defined explicitly in a separate input file, thus also allowing for additional augmentation of basis sets, which is of utmost importance when describing molecular anions.\cite{herbert2015, skurski2000}
The \texttt{handlerG09} and \texttt{handlerQChem} modules provide an interface to the external programs by creating input files and calling the respective programs. The \texttt{dysonG09} and \texttt{dysonQChem} modules contain classes that parse the external output files and organize the data into the form needed in the program.

\subsection{Generation of initial conditions}
\label{init-cond}

The initial nuclear coordinates and velocities are determined by stochastic sampling of an appropriate probability distribution function for the harmonic normal modes of the system. 
These can be computed from the electronic Hessian matrix at an optimized geometry of the studied molecule. 
For molecules in the vibrational ground state as well as for a thermal ensemble of molecules, the Wigner function
\begin{align}
    \rho_W(\{Q_i,P_i\})=\frac{1}{(\pi\hbar)^N}\prod_{i=1}^N \alpha_i(T)\,\mathrm{exp}\left(-\frac{\alpha_i(T)}{\hbar\omega_i}(P_i^2+\omega_i^2Q_i^2)\right)
\end{align}
with
\begin{align}
\alpha_i(T)
=
\tanh{\bigg(\frac{\hbar\omega_i}{2k_BT}\bigg)}
\end{align}
is employed, where $\{Q_i,P_i\}$ denote the normal coordinates and momenta, $\omega_i$ is the angular frequency of normal mode $\nu_i$ and $T$ the thermodynamic temperature.

Besides these cases, in experiments investigating vibration-induced autoionization another type of initial conditions is often important in which one or more normal vibrations of the system are excited by laser irradiation. 
In principle, the respective initial conditions could be also generated by using a Wigner function. However, Wigner functions for excited vibrational states can assume negative values and can thus not be directly identified with a probability distribution. 
A possible approach might be to regard the positive and negative parts of the Wigner function separately as probability distributions and to run a "positive" and a "negative" ensemble of initial conditions, the final properties of the system then being obtained by appropriate averaging. 
As a more efficient alternative, which gets on with only one single ensemble, we employ a positive definite probability distribution constructed from the excited-vibrational state wavefunctions in position and momentum space, 
\begin{align}
\rho^{(i)}_\upsilon(Q_i,P_i)=|\chi^{(i)}_\upsilon(Q_i)|^2|\tilde{\chi}^{(i)}_\upsilon(P_i)|^2,    
\end{align}
where $\chi^{(i)}_\upsilon(Q_i)$ and $\tilde{\chi}^{(i)}_\upsilon(P_i)$ are the harmonic oscillator wavefunctions for quantum state $\upsilon$ of normal mode $\nu_i$ in position and momentum space, respectively.

\subsection{Nuclear dynamics}

Given Newton's equations of motion (\ref{newton}), the nuclei are propagated by numerical solution using the velocity Verlet algorithm \cite{veloverlet} for a user-defined time step.
Within this algorithm, the nuclear coordinates at $t+\Delta t$ are obtained from a Taylor series expansion around the coordinates at $t$:
\begin{align}
\textbf{R}(t + \Delta t) 
\approx
\textbf{R}(t) + 
\dot{\textbf{R}}(t)\Delta t + 
\frac{1}{2} M^{-1}\textbf{F}(t) \Delta t^2,
\end{align}
where in the last term the acceleration has been formulated using the force \textbf{F} given by the electronic potential energy gradient (cf. Eq. (\ref{newton})).
With the new nuclear coordinates, the force at $t + \Delta t$ can be evaluated, giving rise to the new nuclear velocities
\begin{align}
\dot{\textbf{R}}(t + \Delta t)
= 
\dot{\textbf{R}}(t) +
\frac{\Delta t}{2} M^{-1} \left[
    \textbf{F}(t) + \textbf{F}(t + \Delta t)
\right]
.
\end{align}
Due to the approximative nature of the algorithm above and the accuracy of the calculated energy gradients, it is possible that the velocities develop small overall translational or rotational components although the initial conditions were determined with these degrees of freedom set at rest. 
These numerical inaccuracies are detected, in the case of translational velocity by the shift of the center of mass away from the origin of the coordinate system, in the case of rotation by the calculation of the angular velocity according to 
\begin{align}
\boldsymbol{\omega}_{rot} = I^{\,-1}  \textbf{L}
\end{align}
with the moment of inertia $I$ and the angular momentum \textbf{L}. 
The translational and rotational portions of the nuclear velocities are then subtracted from the total velocity and the remaining vibrational part is rescaled to ensure energy conservation.

After each nuclear dynamics step, the new nuclear coordinates and velocities are written into separate output files, the coordinates in a format of consecutive xyz files which can be visualized easily by external software (for example with the VMD program package \cite{vmd}, which is warmly recommended).

\subsection{Electronic dynamics}

Since the evaluation of electronic coupling terms in Eq. (\ref{schroedinger}) is, apart from the external quantum-chemistry calculations, the computationally most expensive step in the dynamics, several approximations need to be implemented, which will be discussed in the following

\subsubsection{Calculation of coupling terms}
\label{subsection:program:couplings}

Before calculating the coupling terms, the discretization procedure for the generation of wave vectors needed to construct the continuum state wavefunctions will be discussed. 
To uniformly discretize angular orientation and kinetic energy of ejected electrons, it is natural to discretize angular and energetic distribution separately. 
Since the kinetic energy of a plane wave is 
\begin{align}
E_{kin}(\textbf{k}_i) = \frac{\hbar^2 |\textbf{k}_i|^2}{2 m_e}
\end{align}
and therefore proportional to the length of the wave vector squared, this length is discretized such that the desired energy range is covered evenly.
For a given energy, the vector orientations are approximately evenly distributed according to the Fibonacci sphere algorithm \cite{fibsphere}.
The volume elements $\Delta {\cal V}_k$ needed for calculating the bound-continuum couplings in Eqs. (\ref{hik}) and (\ref{dik}) are constructed as the difference of spherical caps around the corresponding wave vectors with the base diameter as an average over the six nearest points on the sphere surrounding the vector. 

In the diabatic coupling terms in the AO basis (Eq. (\ref{vdia_ao_main})) two types of four-center integrals are present: (i) such involving four Gaussian-type atomic orbitals (GTOs), $\braket{\sigma \lambda | \mu \nu}$. These are evaluated by using the \texttt{libcint} library \cite{libcint} within the PySCF program package \cite{pyscf1,pyscf2}. 
(ii) integrals involving a plane wave of wave vector $\mathbf{k}_i$ and three GTOs, $\braket{\mathbf{k}_i \lambda | \mu \nu}$. 
These terms can in principle be calculated analytically as outlined, e.g., in Ref. \onlinecite{colle1987}, but this is computationally unfeasible for the present purpose since an immense number of plane waves has to be included for a proper discretization of the ionization continuum. Instead, the plane waves are approximated by their Taylor expansion around the center of basis function $\ket{\mu}$, $\textbf{R}_\mu$. 
As will be discussed in the Performance Section later on, for sufficient accuracy in the approximation it is necessary to include not only the zero'th order term (assuming the plane wave to be constant in the vicinity of the molecule), but also the first-order term, resulting in the approximation
\begin{align}
\mathrm{e}^{i \textbf{k} \cdot \textbf{r}}
&=
\mathrm{e}^{i \textbf{k} \cdot \textbf{R}_\mu}
\mathrm{e}^{i \textbf{k} \cdot (\textbf{r} - \textbf{R}_\mu)}\nonumber\\
&\approx
\mathrm{e}^{i \textbf{k} \cdot \textbf{R}_\mu}
\left[
    1 + i \textbf{k} \cdot (\textbf{r} - \textbf{R}_\mu)
\right].
\end{align}
This leads to two terms for the two-electron integrals as follows:
\begin{align}
\braket{\mathbf{k}_i \lambda | \mu \nu}
\approx
\mathrm{e}^{i \textbf{k} \cdot \textbf{R}_\mu}
\left[
    \braket{\lambda | \mu \nu}
    +
    i \textbf{k} 
    \braket{\lambda | \tilde{\mu} \nu}
\right].
\label{4centerapprox}
\end{align}
In the above expression, $\ket{\tilde{\mu}}$ is an AO basis function with an angular momentum quantum number by one higher than $\ket{\mu}$ while having the same Gaussian exponent. 
This heavily reduces the amount of two-electron integrals to be computed from $n_{AO}^3 n_{PW}$ to $n_{AO}^2 [n_{AO} + n'_{AO}]$, with $n_{AO}$ being the total number of AO basis functions, $n'_{AO}$ the total number of basis functions with increased quantum number and $n_{PW}$ the total number of plane waves. For instance, in the case of vinylidene in Ref. \onlinecite{aid}, this amounts to a reduction by a factor of $\sim$30000.
These terms are again evaluated using the PySCF module.
The prefactors $A$, $\bar{A}$ and $B$ present in Eq. (\ref{vdia_ao_main}) are straightforwardly implemented in Python according to Eqs. (\ref{Afactor}), (\ref{Abarfactor}) and (\ref{Bfactor}). 
Evaluation of the Dyson orbitals needed for the calculation of the nonadiabatic couplings is implemented as described before in Ref. \onlinecite{humeniuk2013} for arbitrary basis sets for the anion and the neutral molecule.
After construction of the Dyson orbitals from all bound anionic states to the neutral ground state the nonadiabatic coupling terms are then calculated according to Eq. (\ref{nonad_pw_mo}). To ensure that the wavefunctions of bound states do not switch their arbitrary signs (which can happen, since the external quantum-chemistry calculations are independent of each other), the overlap of electronic wavefunctions of all bound states are tracked throughout the trajectories and accounted for in all formulae involving the respective states.

\subsubsection{Calculation of electronic state coefficients}
\label{subsection:program:elstatecoef}

The electronic degrees of freedom are propagated by solving the time-dependent Schrödinger equation (\ref{schroedinger}) in the manifold of all considered bound anion and continuum electronic states using Adams' method as implemented in the \texttt{ode} class of Python's \texttt{scipy.integrate} module \cite{scipy} with a user-defined integration time step. For increased computational stability the equations are beforehand transformed into the interaction picture, introducing the new electronic state coefficients
\begin{align}
a_n(t) 
= 
c_n(t)\ 
\mathrm{e}^{\frac{i}{\hbar} H_{nn} t}.
\label{interaction1}
\end{align}
Inserting this into Eq. (\ref{schroedinger}) leads to the actually implemented electronic equation of motion
\begin{align}
\dot{a}_n(t)
=
\sum_m 
    \left[ 
        -\frac{i}{\hbar} \tilde{H}_{nm} - D_{nm}
    \right] 
    a_m(t)
    \mathrm{e}^{-\frac{i}{\hbar} (H_{mm} - H_{nn}) t}
\end{align}
where $\tilde{H}_{nm}$ denotes the Hamiltonian matrix of the system with zeros on the diagonal.

\subsubsection{Hopping procedure}

Hopping probabilities are directly evaluated according to Eq. (\ref{probs}) from the state coefficients: A random number between 0 and 1 is generated using the \texttt{random} function in the \texttt{numpy.random} module and hopping is conducted accordingly.
Once a trajectory hops into a continuum state, it could in principle be straightforwardly continued on the potential energy surface of the neutral molecule.
Although this can be quite insightful if one is interested in the subsequent geometric changes of the ionized system, we follow a different approach and stop the trajectories after electron detachment since our focus is set on the actual autoionization process.
This allows us to implement a modification of the surface-hopping procedure that leads to a great improvement of the hopping statistics. The idea is to divide a single trajectory into 'sub-trajectories', i.e. to evaluate if a trajectory hops a number $n_{subtraj}$ of times (see Fig. \ref{fgr:flowchart}). 
Every time a sub-trajectory hops into the continuum, $n_{subtraj}$ is reduced by one and once it reaches zero, the underlying nuclear dynamics is stopped. 
It has to be noted that this procedure is only followed for hops into the continuum, while for hops between bound anionic states only a single hopping event per trajectory and time step is possible due to the need to continue the nuclear dynamics on an unambiguously determined potential energy surface.

\subsection{Graphical user interface}

\begin{figure*}[tb]
 \centering
 \includegraphics[width=\textwidth, draft=false]{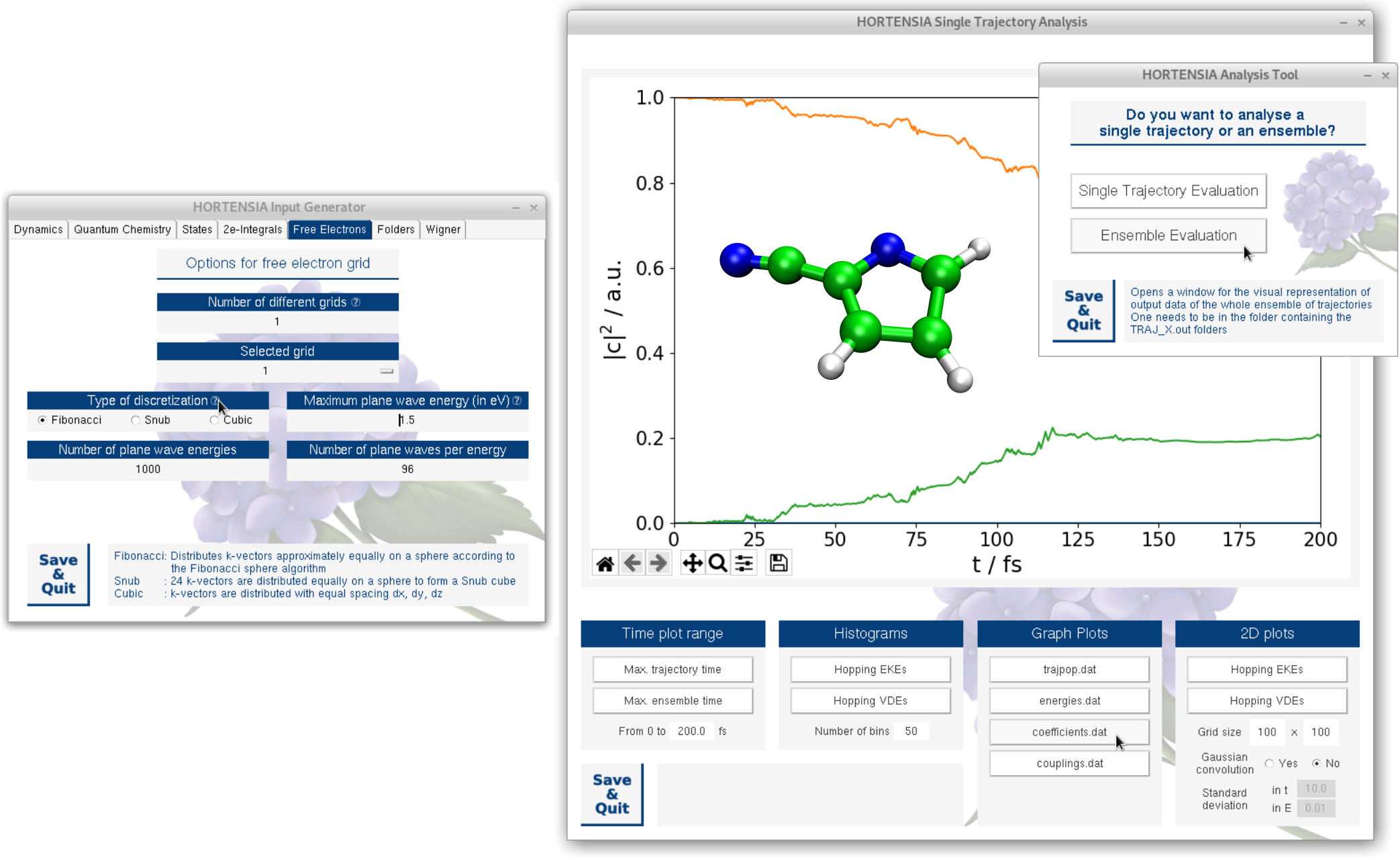}
 \caption{Left: example page of the graphical input generation tool; right: output analysis GUI with the electronic state coefficients of a single, exemplary 2-cyanopyrrolide trajectory. The molecular structure is only an illustrative image, which was created using the VMD program, and its creation is not part of the presented program.}
 \label{fgr:gui}
\end{figure*}

Our program package comes with a graphical user interface (GUI) for the input generation as well as an analysis tool for trajectories. 
An example of the former is displayed in Fig. \ref{fgr:gui}. 
In the input generator, which is started with 
\begin{lstlisting}[language=bash]
  $ hortensia --gui
\end{lstlisting}
in addition to all relevant settings for the actual simulation, the user may find options for the generation of a complete folder structure for the trajectories as well as bash submit scripts to be used with the Slurm Workload Manager\cite{slurm}. 
Furthermore, the above mentioned Wigner ensemble scripts can be used and initial conditions can be generated. Therefore it is highly recommended to use the GUI feature.

Additionally, through the command
\begin{lstlisting}
  $ hortensia --analysis
\end{lstlisting}
one can open the analysis tool which is able to read output files and visualize them in a sub-window using the \texttt{matplotlib} program package \cite{matplotlib}.

\subsection{Installation}

The most convenient way to install the program package is downloading or cloning the \href{https://github.com/mitric-lab/HORTENSIA_LATEST.git}{\textit{repository on our Github page}}\cite{hortensia}. In the main folder, execute
\begin{lstlisting}[language=bash]
  $ python cysetup.py build_ext --inplace
  $ pip install .
\end{lstlisting}
to first compile the Cython modules and then install the program. The program package requires (and will automatically pip install) 
\begin{itemize}
    \item \texttt{python >= 3.8}
    \item \texttt{cython} - for faster summation of large arrays, mainly in the calculation of the two-center integrals in Eqs. (\ref{vdia_ao_main}) and (\ref{4centerapprox})
    \item \texttt{scipy} - mainly in the integration of the electronic Schrödinger equation as outlined in subsection \ref{subsection:program:elstatecoef}
    \item \texttt{pyscf} - for the calculation of the two-electron integrals in Eqs. (\ref{vdia_ao_main}) and (\ref{4centerapprox})
    \item \texttt{joblib} - for the parallelization of diabatic couplings
    \item \texttt{matplotlib} - for the plots in the sub-window of the analysis tool described before
\end{itemize}
and all dependencies thereof.
Using the command
\begin{lstlisting}[language=bash]
  $ pip uninstall hortensia_latest
\end{lstlisting}
will uninstall the program package.

\section{Discussion}
\label{section:discussion}

In this section we will quantify aspects of the program related to overall performance. 
This includes the quality of approximations within the methodology as well as optimization of time consumption and computational resources.
Moreover the exemplary autoionization dynamics of the 2-cyanopyrrolide anion is discussed.

\subsection{Accuracy of \textbf{k}-vector discretization and integral approximations}

\begin{figure}[tb]
 \centering
 \includegraphics[width=\columnwidth, draft=false]{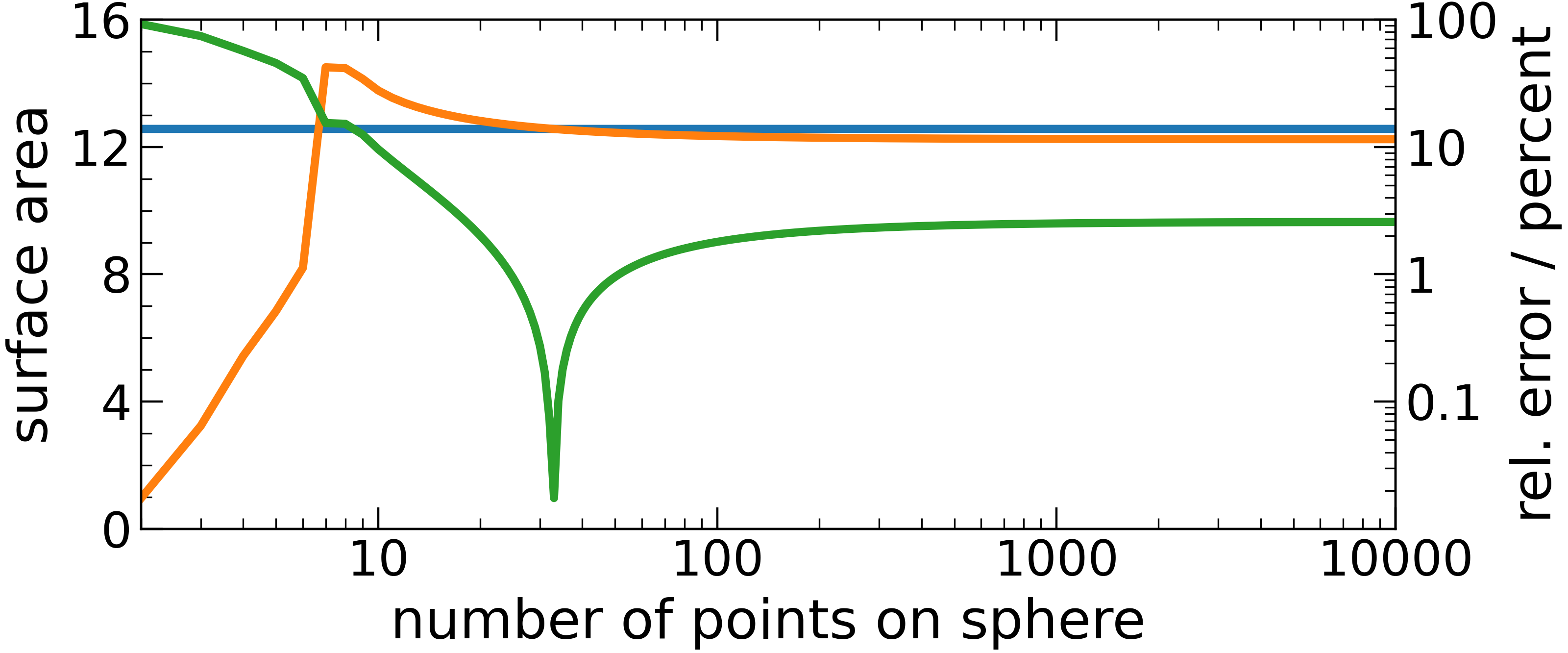}
 \caption{Comparison of the actual surface area of a unit sphere ($A_{\mathrm{sphere}} = 4 \pi$, blue line) and the approximated surface area as described in subsection \ref{subsection:program:couplings} for up to 10$^4$ vector orientations (orange). The relative error is given in green.}
 \label{fgr:fibError}
\end{figure}

The accuracy of the Fibonacci sphere algorithm for angular discretization in \textit{k}-space is illustrated in Fig. \ref{fgr:fibError} by the covered surface area of a unit sphere using a given number of distributed points. 
The total surface area (orange graph) is presented with the relative error $|A_{\mathrm{fib}}-A_{\mathrm{sphere}}|/A_{\mathrm{sphere}}$ (green graph) to the exact surface area  $4\pi \approx 12.566$ (blue line). 
The approximated area rapidly converges to a value of $\sim$12.243, which corresponds to a relative error of $\sim$2.575 \%.
Since in the coverage of k-vector lengths no additional approximation is introduced and for their respective volume elements the k-space is divided energetically evenly (thus covered exactly with respect to vector length), the error in the surface area for specific vector lengths equates to the overall error of the volume elements.
Therefore the sum of these volume elements results in a total volume that deviates by less than 3 \% from the actual sphere for arbitrary numbers of vector orientations $n_s \ge 30$ and lengths $n_{E}$ (giving a total number of wave vectors $n_k = n_E \cdot n_s$).

\begin{figure}[tb]
 \centering
 \includegraphics[width=\columnwidth, draft=false]{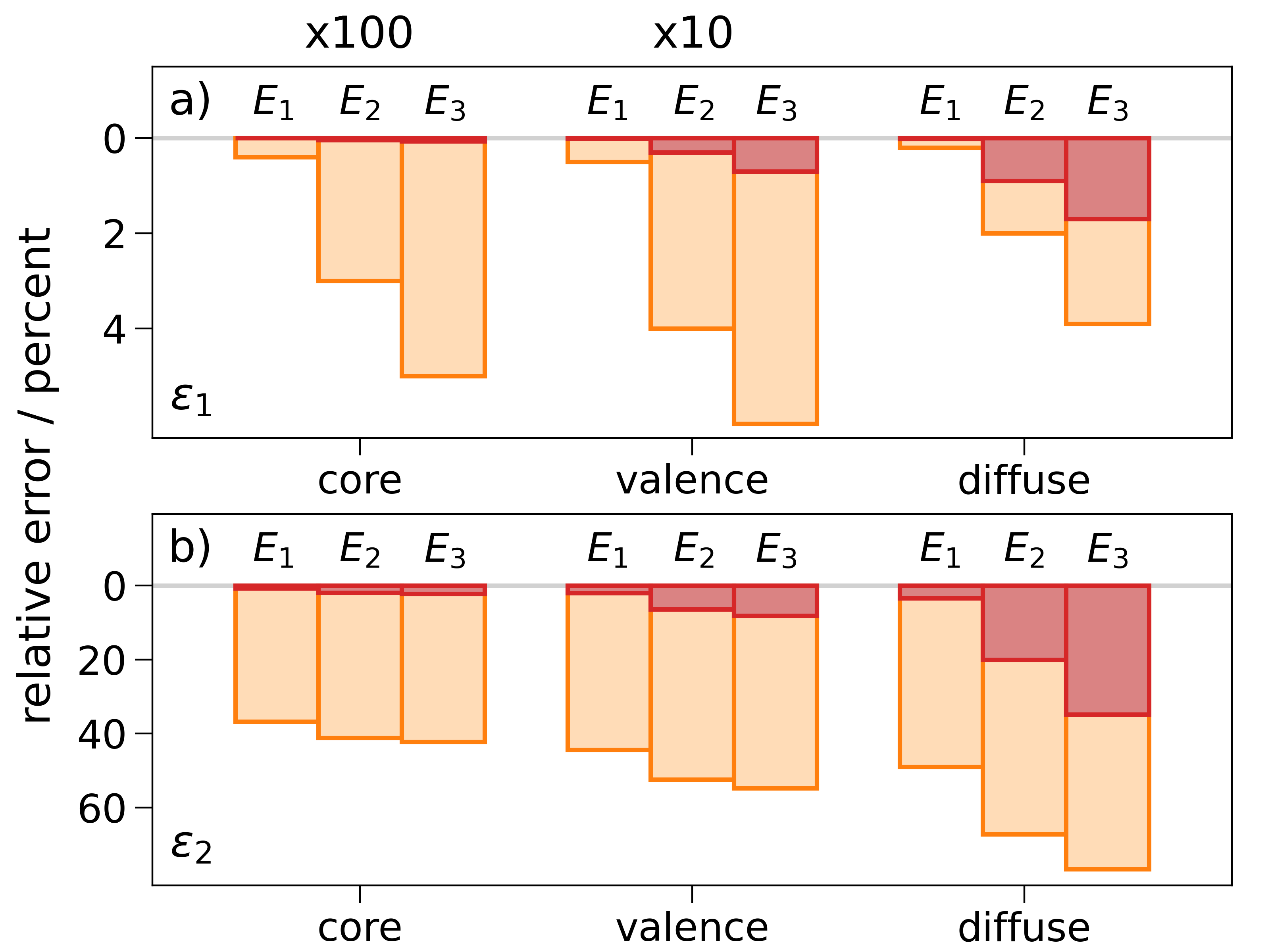}
 \caption{
Errors (in \%) of hybrid Gaussian-plane wave electron repulsion integrals $\langle \mathbf{k}_i\lambda|\mu\nu\rangle$ for 2-cyanopyrrolide employing the 6-311++G** +3s2p basis set. The molecular structure has been optimized in the dipole-bound excited state at the TDDFT/$\omega$B97XD level using the same basis set. 
Two types of error measures are reported: a) $\epsilon_1^{\textrm{ap}}=\braket{|I_{ex}-I_{ap}|}/\braket{|I_{ex}|}$ and b) $\epsilon_2^{\textrm{ap}}~=~\braket{|I_{ex}-I_{ap}|\,/\,|I_{ex}|}^a$, where $I_{ex}$ denotes the exact integral, $I_{ap}$ the approximate value either according to Eq. (\ref{4centerapprox}) (linear, red bars) or assuming the plane wave to be constant (orange bars). For $\epsilon_2^{\textrm{ap}}$ the average  has been computed for all integrals with $|I_{ex}|>10^{-16}\,E_H$.
To compute the average the integrals are grouped according to the exponent $\alpha$ of basis function $\mu$ as core ($\alpha>10\,a_0^{-2}$), valence ($10\,a_0^{-2}<\alpha<0.1\,a_0^{-2}$), and diffuse ($\alpha<0.1\,a_0^{-2}$). 
For each plane wave energy ($E_1 = 0.0015$ eV, $E_2 = 0.1$ eV, $E_3 = 0.2$ eV), the average has been taken over all distinct integrals provided by the basis set as well as over 24 different k-vectors corresponding to the direction vectors of the vertices of a snub cube.
In a), the core and valence error bars are multiplied by a factor of 100 and 10, respectively, to enhance visibility.
}
 \label{fgr:4c_comparison}
\end{figure}

The approximation of the plane wave by the first terms of its Taylor expansion as introduced in Eq. (\ref{4centerapprox}) relies on the assumption that the amplitude of the plane wave only changes marginally within the extent of the AOs. 
Fig. \ref{fgr:4c_comparison} shows a comparison between the approximation with linear terms, an even simpler constant-wave approximation where $\textrm{e}^{i\textbf{k}\textbf{r}} \approx \textrm{e}^{i\textbf{k}\textbf{R}_\mu}$ and the exact integrals for selected plane wave vectors for 2-cyanopyrrolide, which serves as an example molecule illustrating the applicability of the program (see section \ref{cyapide} below). 
Two error measures are compared: a relative value of average deviations ($\epsilon_1$) in Fig. \ref{fgr:4c_comparison}a) and an average value of relative deviations ($\epsilon_2$) in Fig. \ref{fgr:4c_comparison}b), which differ insofar as in $\epsilon_1$, the deviations between exact and approximate integrals are averaged first and then divided by the overall average value of the exact integrals, while in $\epsilon_2$, first for each individual integral the relative error is computed, followed by averaging the results. 
The averages are reported for three illustrative plane wave energies and grouped according to the Gaussian exponent of the basis function sharing its electron coordinate with the plane wave as "core", "valence" and "diffuse" with decreasing size of the exponent (for details see Fig. \ref{fgr:4c_comparison}). Overall, it becomes evident that for both error measures, the linear approximation of the plane wave is clearly superior to the constant approximation. The values of $\epsilon_1$ are always much smaller than those of $\epsilon_2$, which is due to the fact that the relative errors of smaller integrals tend to be larger than those of larger ones, and the definition of $\epsilon_1$ partially compensates for this fact. Errors larger than a few percent only occur for $\epsilon_2$ calculated for diffuse basis functions at larger plane wave energies. Since in the actual computations, the approximate integrals are employed to calculate the diabatic couplings and for this, the sum over the entire basis set is taken (cf. Eq. (\ref{vdia_ao_main})), especially the smallness of error $\epsilon_1$ encourages the use of the linear approximation.

\subsection{Optimization of program performance}
Where computationally advantageous, we separate the time-dependent and -independent parts of the underlying equations and pre-calculate the time-independent terms at the beginning of the simulation. 
This results in higher overall memory usage, however of only several hundred MB to a few GB (depending on the molecular system), but leads to significant time-saving, which is still a desirable trade-off when calculating on CPU clusters but may limit the use of the program on single desktop computers.

Furthermore, for increased performance the summation over the four-center integrals in terms 2 and 4 on the right side of Eq. (\ref{vdia_ao_main}) is implemented as follows: one first pre-calculates the terms $A$, $\bar{A}$ and $B$ given in Eqs. (\ref{Afactor})-(\ref{Bfactor}) for all AOs. 
Then the calculation of the four-center integrals using the PySCF program package is divided into $n_{proc}$ smaller terms, $n_{proc}$ being the user-defined number of processors, and then evaluated in parallel utilizing the \texttt{joblib}\cite{joblib} library by explicit summation over all AO combinations implemented in a Cython\cite{cython} module, therefore reducing the memory usage by ridding oneself from massive arrays while also improving the runtime performance of this time bottleneck through parallelization.

Together with the calculation of coupling terms, the most time-consuming step of the simulation is the two external quantum-chemical calculations needed in each time step. 
There are a few options to improve the performance of these calculations, the easiest of which are to increase the number of utilized processors and to reduce convergence time by loading the results of the last time step as an initial guess for the new calculation. 
Another possibility is in the choice of basis sets.
Finding a basis set for anions prone to autoionization can be challenging due to the small ionization energies and the diffusity of the states that comes with it.
Therefore one has to consider basis sets augmented with enough diffuse basis functions to reasonably describe the properties of the system \cite{herbert2015}.
Although popular basis sets such as doubly and triply augmented Dunning-style basis sets (daug-cc-pVDZ, taug-cc-pVDZ) are (generally speaking) a potentially good choice for the description of loosely-bound anions, the size of these basis sets is computationally prohibitive if one aims to run dynamics simulations and therefore thousands of consecutive quantum-chemical calculations. 
A good alternative can be the usage of smaller basis sets (such as 6-311++G**\cite{631pGss, 6311ppGss}) augmented with additional diffuse functions generated by geometric progressions of the Gaussian exponents as outlined in Ref. \onlinecite{skurski2000}.

Considering the overall time consumption, no real performance benchmarks exist with which to compare our program package, since the theory behind it is rather novel. 
Therefore we will briefly discuss the specific case of vinylidene from our work presented in Ref. \onlinecite{aid} and the 2-cyanopyrrolide example discussed in detail in section \ref{cyapide} below.

The vinylidene dynamics was performed for a total time of 3 ps in 15000 nuclear dynamics time steps at the $\omega$B97XD\cite{wb97xd}/d-aug-cc-pVDZ level of theory, which consists of 146 primitive Gaussian basis functions and 96000 plane waves, amounting to $\sim$460 million 2-electron integrals per time step to be solved (cf. Eqs. (\ref{vdia_ao_main}) and (\ref{4centerapprox})). 
Using 6 Intel Xeon E5-2660 (v3) processors per trajectory, the average computation time was around 11 days and 14 hours with a peak memory usage of $\sim$9 GB. Of the total time, around 5 days (or 43 \%) were needed for the external quantum-chemistry calculations with the Gaussian09 program package. 
It also has to be noted that of the remaining time $\sim$30 \% can be attributed to the calculation of the 2-electron integrals in the diabatic couplings running on a single processor, which has since been parallelized for improved performance.

In the simulation of the 2-cyanopyrrolide dynamics, the 6-311++G**+3s2p basis set consists of 297 primitive Gaussian basis functions which results in $\sim$7.8 billion 2-electron integrals to be summed over per time step.
The average computation time amounted to 13 days and 12 hours on 10 Intel Xeon E5-2660 (v3) processors per trajectory, for a total time of 200 fs in 1000 nuclear time steps. 
The inclusion of an excited state leads to a massive increase in time consumption in the quantum-chemical calculations (which account for $\sim$47 \%/ 6.4 days of the total computation time) as well as the evaluation of diabatic couplings, where the summation of all integral terms (cf. Eq. (\ref{vdia_ao_main})) is now also conducted on 10 processors using the \texttt{joblib} module.

\begin{table*}[tb]
\caption{\label{tab:cyapide}
Comparison of adiabatic electron affinities (AEA), vertical detachment energies to the neutral ground state (VDE$_\textrm{GS}$), vertical attachment energies to the anionic ground state (VAE$_\textrm{GS}$) and excitation energy to the dipole-bound state ($\Delta$E$_{\textrm{DBS}}$) for the 2-cyanopyrrolide anion. The superscript indicates at which optimized geometry (a = anion, n = neutral) the respective value is calculated. The method used in the dynamics simulation is indicated in bold font. The added basis functions +Xs etc. are generated according to Ref. \onlinecite{skurski2000} (with a factor for the geometric progression of 3.5) and centered on the nitrogen atoms. All energies values are given in eV.
In some cases AEA$_{\textrm{DBS}}$ could not be obtained, instead the vertical attachment energy at the neutral equilibrium geometry is given as an approximation (denoted with *).
$^\textrm{a}$ Ref. \onlinecite{augccpvxz1},
$^\textrm{b}$ Ref. \onlinecite{augccpvxz1, augccpvxz2},
$^\textrm{c}$ Ref. \onlinecite{augccpvxz1, augccpvxz2, xaugccpvxz},
$^\textrm{d}$ Ref. \onlinecite{631pGss},
$^\textrm{e}$ Ref. \onlinecite{631pGss, 6311ppGss},
$^\textrm{f}$ Ref. \onlinecite{ccsd1, ccsd2}
}
\begin{ruledtabular}
\begin{tabular}{lccccc}
Method & AEA$_\textrm{GS}$ & AEA$_\textrm{DBS}$ & VDE$^\textrm{a}_\textrm{GS}$ & VAE$^\textrm{n}_\textrm{GS}$ & $\Delta$E$^\textrm{n}_{\textrm{DBS}}$\\[1ex]
\hline
$\omega$B97XD / aug-cc-pVDZ$^\textrm{b}$   & 3.075 & -0.674  & 3.225 & 2.932 & 3.612 \\
$\omega$B97XD / d-aug-cc-pVDZ$^\textrm{c}$ & 3.071 & -0.117* & 3.221 & 2.929 & 3.046 \\
$\omega$B97XD / t-aug-cc-pVDZ$^\textrm{c}$ & 3.070 &  0.031* & 3.220 & 2.928 & 2.897 \\
$\omega$B97XD / t-aug-cc-pVTZ$^\textrm{c}$ & 3.044 & 0.044* & 3.200 & 2.899 & 2.855 \\
$\omega$B97XD / 6-31+G**$^\textrm{d}$      & 3.062 & -1.374  & 3.212 & 2.919 & 4.302 \\
$\omega$B97XD / 6-31+G** +3s2p             & 3.064 &  0.060* & 3.217 & 2.921 & 2.861 \\
$\omega$B97XD / 6-311++G**$^\textrm{e}$    & 3.095 & -0.759  & 3.249 & 2.949 & 3.716 \\
$\omega$B97XD / 6-311++G** +3s             & 3.095 &  0.064  & 3.249 & 2.949 & 2.887 \\
\textbf{$\omega$B97XD / 6-311++G** +3s2p}  & \textbf{3.094} &  \textbf{0.063}  & \textbf{3.248} & \textbf{2.948} & \textbf{2.887} \\
$\omega$B97XD / 6-311++G** +3s2p2d         & 3.076 &  0.055* & 3.230 & 2.930 & 2.875 \\
\hline
EOM-CCSD$^\textrm{f}$ / t-aug-cc-pVDZ      & 2.844 & -0.102* & 3.016 & 2.686 & 2.788 \\
EOM-CCSD / 6-31+G** +3s2p                  & 2.679 & -0.346* & 2.855 & 2.518 & 2.864 \\
EOM-CCSD / 6-311++G** +3s2p                & 2.748 & -0.148* & 2.929 & 2.584 & 2.732 \\
\hline
Experiment\cite{wang2022}                  & 3.0981 & 0.0298 &       &       &  \\
\hline
\end{tabular}
\end{ruledtabular}
\end{table*}

\begin{figure}[tb]
 \centering
 \includegraphics[width=\columnwidth, draft=false]{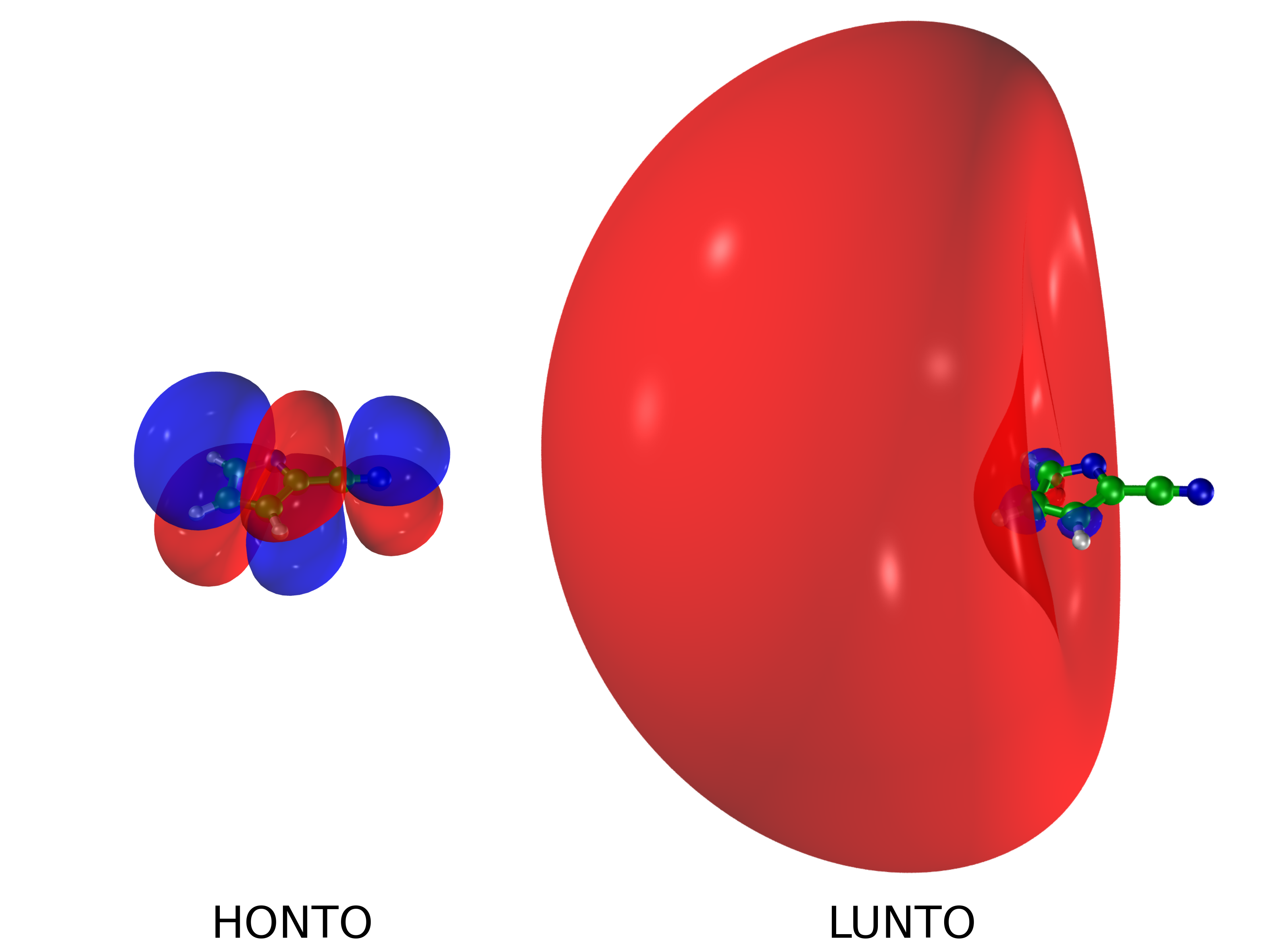}
 \caption{HONTO and LUNTO of 2-cyanopyrrolide at the optimized geometry of the dipole-bound first excited state at the $\omega$B97XD/ 6-311++G** + 3s2p level of theory with an isovalue of 0.003.}
 \label{fgr:NTOs}
\end{figure}

\begin{figure}[tb]
 \centering
 \includegraphics[width=\columnwidth, draft=false]{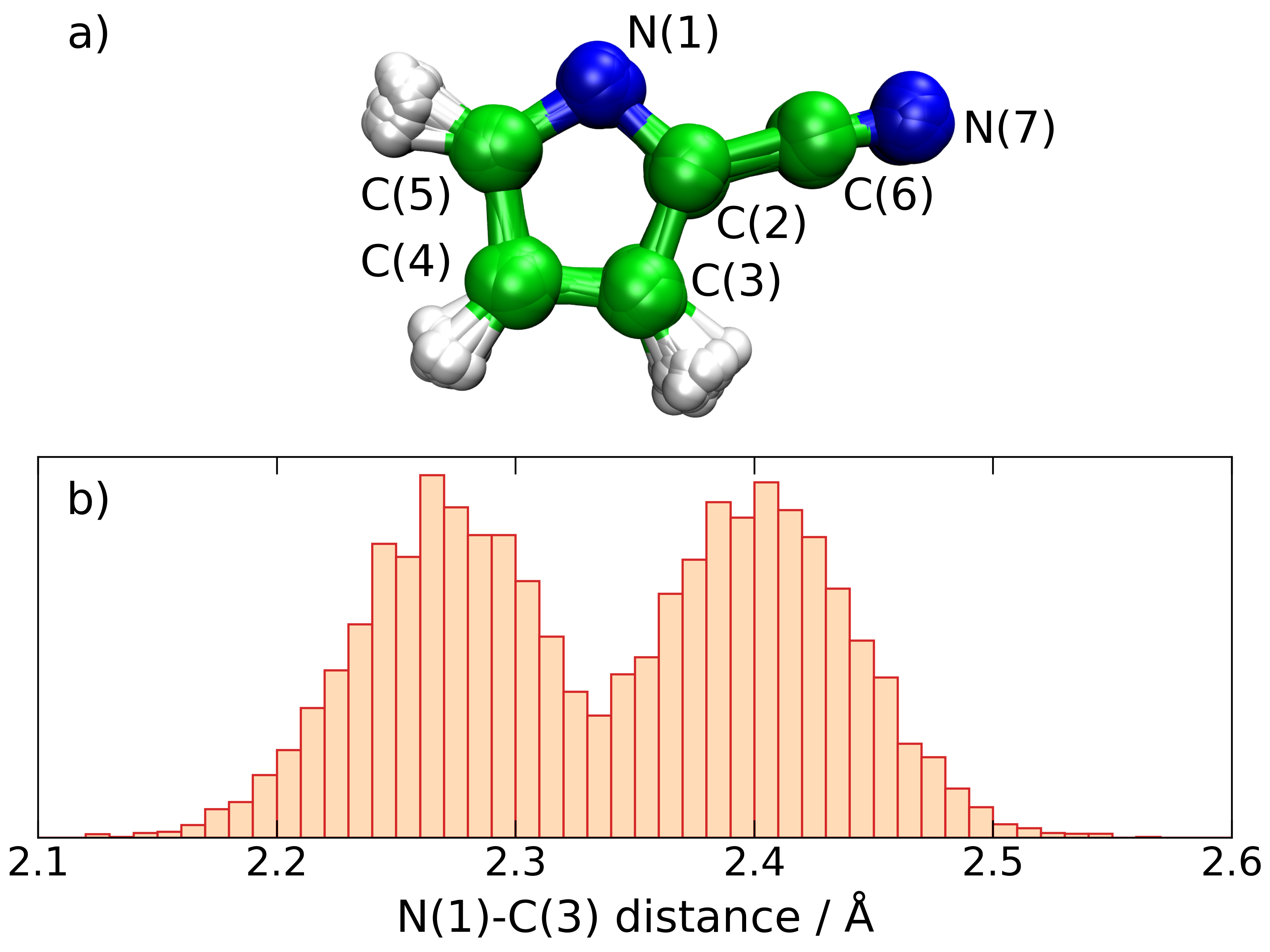}
 \caption{a) Overlay of all initial molecular structures used in the dynamics simulation of 2-cyanopyrrolide, b) distribution for 10000 initial conditions as a function of distance (in \AA) between the nitrogen (1) and carbon (3) atom as marked in a), showing a bimodal structure.}
 \label{fgr:initialCond}
\end{figure}

\begin{figure}[tb]
 \centering
 \includegraphics[width=\columnwidth, draft=false]{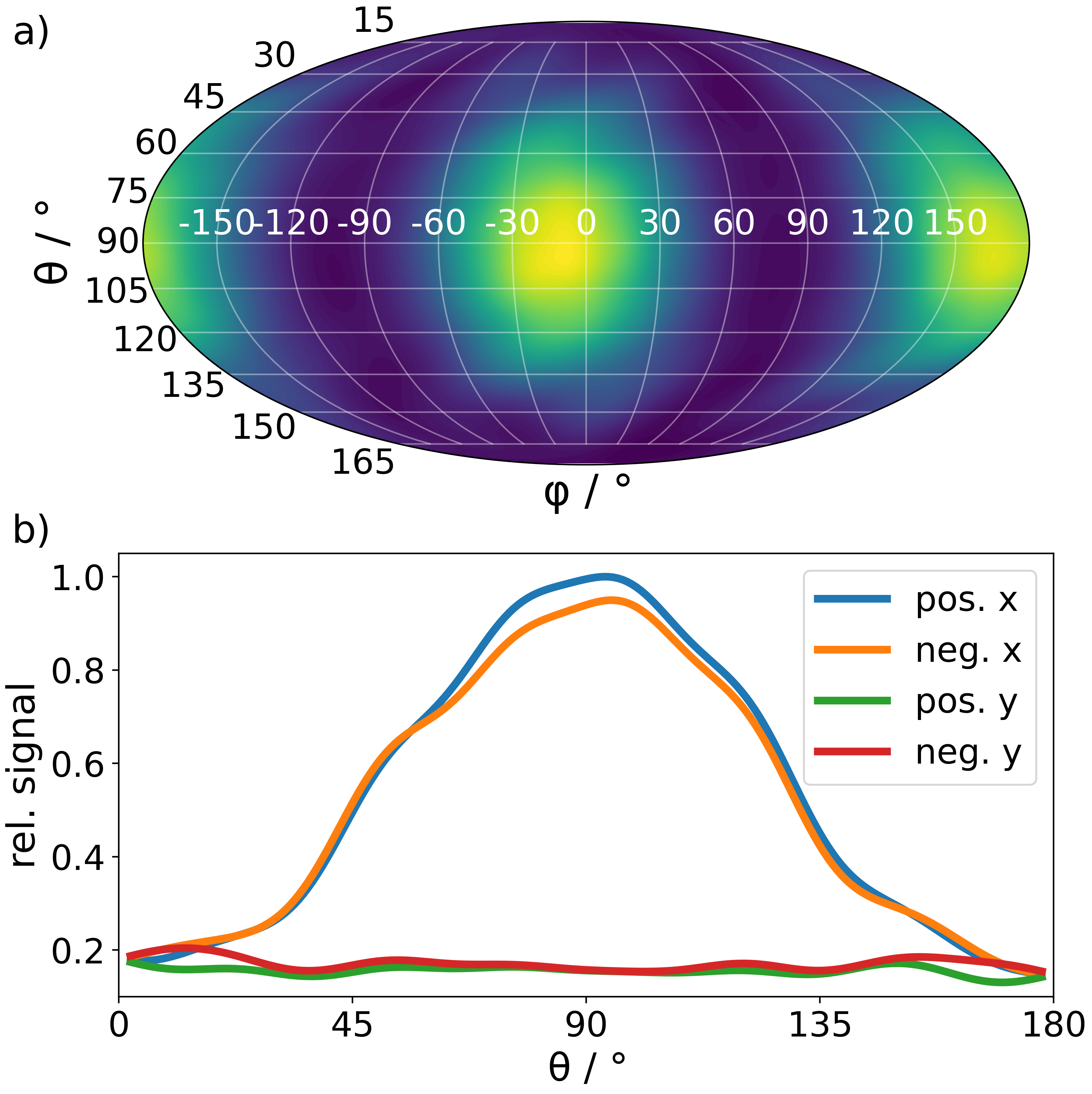}
 \caption{a) Mollweide projection of the angular distribution of ejected electrons in the 2-cyanopyrrolide dynamics, summed over all energies. The x-axis ($\varphi = 0$, $\theta = 90$ degrees) is aligned with the cyano group and the molecule lies within the xy-plane ($\theta = 90$ degrees); b) Slices through the Mollweide projection at $\varphi$ angles of 0 (positive x direction, blue), 180 (negative x direction, orange), 90 (positive y direction, green) and 270 (negative y direction, red) degrees.}
 \label{fgr:angle2d}
\end{figure}

\begin{figure}[tb]
 \centering
 \includegraphics[width=\columnwidth, draft=false]{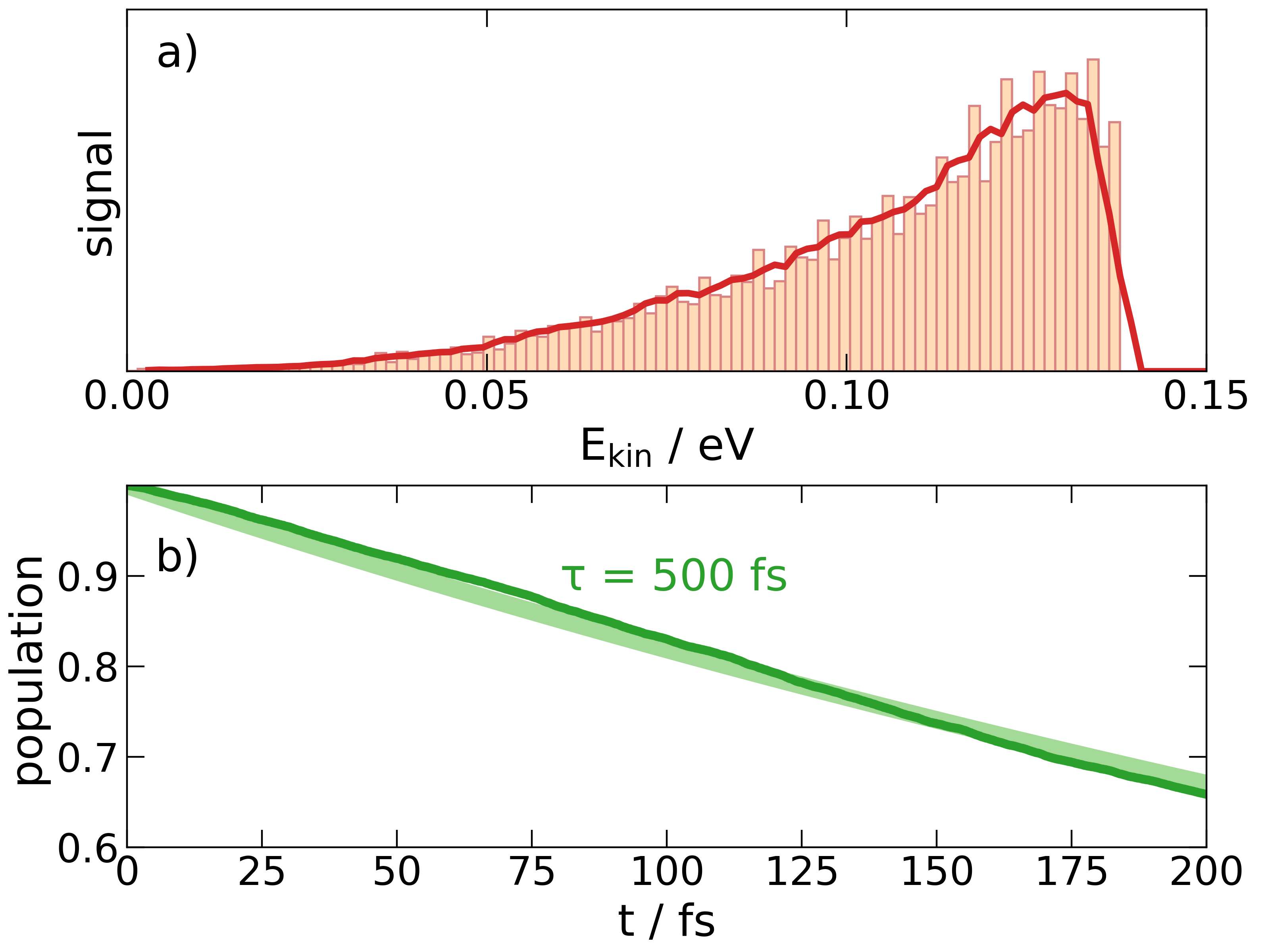}
 \caption{a) Simulated electron kinetic energy distribution of all hopping events after excitation of mode $\nu_{11}$ and propagation for 200 fs (orange histogram) and running average over 5 points/7.5 meV (red curve), b) time-dependent population of all bound anion states (dark green) and exponential fit with a time constant of $\tau = 500$ fs (light green).}
 \label{fgr:ekePopCyapide}
\end{figure}

\subsection{Illustrative example: Autoionization of the 2-cyano\-pyrrolide anion}
\label{cyapide}

To illustrate the scope of our program, we simulated the vibration-induced autoionization dynamics of the example anion 2-cyanopyrrolide.
Experimentally, this molecule was measured to have an adiabatic electron affinity of 3.0981 eV and possesses a Rydberg-s type dipole-bound state 29.8 meV below the ionization threshold.\cite{wang2022} As can be seen in Table \ref{tab:cyapide}, which compares several quantum-chemistry methods and basis sets with the data measured by Wang $et\ al.$, the experimental data is reproduced quite well using the $\omega$B97XD functional and large, diffuse basis sets such as triply augmented pVDZ/pVTZ. 
Moreover, although the description of the molecule with standard Pople-type basis sets is fairly inaccurate, further augmentation with extra diffuse basis functions (see Ref. \onlinecite{skurski2000}), in this case placed on the nitrogen atoms, also leads to good agreement with the experimental values. 
At the same time this approach retains a significantly smaller total number of basis functions, therefore keeping computational effort manageable.
Fig. \ref{fgr:NTOs} shows the HONTO and LUNTO, visualizing the spatial distribution of the excess electron in the ground and excited state at the optimized geometry of the dipole-bound first excited state employing the $\omega$B97XD functional and the 6-311++G** basis set augmented with three diffuse s- and two diffuse p-functions on each nitrogen atom (henceforth abbreviated as 6-311++G** + 3s2p). 
The shape of the excess electron's probability distribution in the dipole-bound state is of s-type, showing that employing additional higher polarization functions (d-/f-type) would lead to no further improvement in the description of the system.
This is in complete agreement with a dipole moment of the neutral species of 5.02~D, well below the second critical dipole moment of $\sim$10~D needed for the binding of an electron in a p-type orbital,\cite{jordan2003} consequently resulting in an s-type distribution centered around the positive end of the molecular dipole vector.

Using the 6-311++G** + 3s2p basis set with the $\omega$B97XD functional, we simulated the vibration-induced autoionization dynamics in the first excited state with the normal mode at 946~cm$^{-1}$ of A' symmetry ($\nu_{11}$ when sorted by increasing mode energy irrespective of symmetry) excited by one vibrational quantum. The initial conditions were generated as described in subsection \ref{init-cond}. Mode $\nu_{11}$ involves a symmetric stretching of the C-H bonds at carbon atoms 4 and 5 as well as a ring breathing motion affecting mostly the ring N and carbon 3. The numbering of atoms is provided in Fig. \ref{fgr:initialCond}a), which illustrates the resulting set of initial conditions by the superposition of all initial structures. In Fig. \ref{fgr:initialCond}b), the distance between the ring nitrogen and the carbon 3 is depicted, which exhibits a bimodal distribution typical for an excited vibrational state. The particular choice of vibrational excitation corresponds to the experimentally observed resonance 7 of the photodetachment spectrum in Ref. \onlinecite{wang2022}. 
The simulation was carried out propagating an ensemble of 53 trajectories for a total of 200 fs (1000 nuclear time steps) with a discretized continuum of 400 plane wave energies evenly spaced from 0.0~eV to 0.138 eV and 96 orientations per energy.
The maximum allowed kinetic energy of the plane wave is the sum of the vibrational excitation energy and the difference in zero-point energies of anion and neutral system, that is, the maximum excess energy available upon ionization.

Notice that due to the very low electron binding energy of the dipole-bound state and the approximative nature of the quantum chemically determined energies, it is challenging to precisely reproduce subtle binding energy differences on the meV scale along the trajectories. 
Thus, some instances of negative VDE occur in the dynamics.
However, the experimental data from Ref. \onlinecite{wang2022} only feature a peak attributed to vibrational autoionization.
Therefore, we only include the latter in our simulation and neglect adiabatic ionization.

The nuclear dynamics following the vibrational excitation is characterized by relatively small amplitude motion. This is due to the overall low internal energy of the molecule and its rigidity as a cyclic system. In the course of the dynamics, the molecular dipole moment associated with the neutral core, which is responsible for electron binding in the excited state, exhibits slight oscillatory behavior while being approximately situated in the molecular plane. This leads to an anisotropic ejection of electrons predominantly in the molecular plane along the axis containing the cyano group, as can be inferred from the Mollweide projection of the angle-dependent distribution of \textbf{k}-vectors, summed over all k values shown in Fig. \ref{fgr:angle2d}a). The resulting electron distribution is thus p-shaped, with maxima along the x-(cyano group) axis and minima in the yz-plane exhibiting only about 20\% of the maximal intensity, as can be seen in Fig. \ref{fgr:angle2d}b).
This observation is in line with the qualitative considerations of nonadiabatic autoionization from dipole-bound states outlined in Ref. \onlinecite{simons2020}.
No transitions to the anionic ground state are observed in our simulation due to a large energy gap regardless of geometry, therefore the angular electron distribution is solely due to ionization from the s-type dipole-bound state.

Regarding the electron kinetic energies, the distribution displayed in Fig. \ref{fgr:ekePopCyapide}a) is obtained, exhibiting a broad peak near the maximally possible energy of 0.138 eV. This can be attributed to a transition in which the vibrational energy of the excited mode is transferred completely to the outgoing electron, i.e., the vibrational energy of the molecule is reduced by one quantum in line with the propensity rules for vibrational autoionization established by Simons\cite{simons1981}.
Further analysis of the peak shape should be taken with care, since for conceptual reasons vibrational resolution is not within the scope of quantum-classical dynamics.

Besides the spatial and energetic distribution of the ejected electrons, our simulation provides access to the timescale in which the ionization process takes place. Fig. \ref{fgr:ekePopCyapide}b) shows the time-dependent population of the bound anionic states, which exhibits a rapid decay that can be fit to an exponential function with a time constant of 500 fs. This value corresponds to a spectral width of around 70 cm$^{-1}$, which is of comparable size to the observation made in Ref. \onlinecite{wang2022}.

Overall this example calculation shows the applicability and scope of the method in the context of small to medium sized molecular anions, providing a means to gain molecular-level insight into the spatio-temporal dynamics of vibration-induced autoionization processes complementary to experimental measurements.

\section{Conclusion}
\label{section:conclusion}

We have presented the Python program package HORTENSIA (Hopping real-time trajectories for electron-ejection by nonadiabatic self-ionization in anions) for the simulation of vibration-induced autoionization processes in molecular anions. 
The program implements our recently introduced extended surface hopping approach for the quantum-classical description of nonadiabatic autoionization dynamics, where the electronic degrees of freedom are treated quantum-mechanically, while the nuclear motion is represented by classical trajectories.
The electronic states included in the dy\-na\-mics simulation comprise the bound adiabatic anionic states and discretized 'ionized system' states composed of a neutral core and a free electron wave function, between which nonadiabatic transitions are simulated in a stochastical manner from hopping probabilities obtained from changes in electronic state coefficients according to Tully's fewest-switches algorithm.
The time-dependent state coefficients are calculated by solution of the electronic Schrödinger equation containing the nonadiabatic as well as diabatic couplings between the considered electronic states according to our presented methodology.

As shown in the example of 2-cyanopyrrolide, time- and angle-resolved electron kinetic energy signals are obtained  directly from the surface-hopping trajectories.
Since no deactivation to the ground state is observed in our simulation, autoionization with a time constant of 500~fs is identified as the only available deactivation pathway in the dipole-bound state of 2-cyanopyrrolide on the simulated timescale, with an anisotropic, p-like ejection of electrons along the cyano-axis.
Moreover, with our program geometric data is yielded which allows for the structural analysis of molecules throughout the autoionization dynamics, providing easy access to geometric characteristics of the considered system, as demonstrated extensively in the example of the vinylidene\cite{aid} and 1-nitropropane\cite{issler2023_3} anions.

Furthermore, the implementation and internal structure of our program package was discussed, which also consists of secondary functionalities such as an input generator and a routine for the creation of initial conditions for nuclear coordinates and velocities within an easy-to-operate graphical user interface (GUI). 
Moreover, the program package provides the user with an additional GUI for the analysis and graphical representation of the most important dynamics results.

In the future, useful extensions of the methodology could be the implementation of neutral molecules to be ionized, which requires the description of scattering states interacting with a cationic core, as well as the inclusion of laser field coupling (analogous to the FISH method\cite{mitric2009}) to describe photoionization beyond the perturbative limit, thereby providing an extension of the approach developed in Ref. \onlinecite{humeniuk2013}.
In addition, the treatment of electronically adiabatic autoionization could be combined with an ab inito computation of the electronic resonance lifetimes, e.g., along the lines presented in Ref. \onlinecite{jagau2022}.

\section*{Data availability}
The data that support the findings of this study are available from the corresponding author upon reasonable request.


\section*{Author declarations}
\subsection*{Conflict of Interest}
The authors have no conflicts to disclose.

\subsection*{Author Contributions}
\noindent
Kevin Issler: Data curation (lead); Formal analysis (lead); Investigation (lead); Methodology (equal); Software (lead); Visualization (lead); Writing - original draft (equal).
Roland Mitrić: Conceptualization (equal); Funding acquisition (lead); Methodology (supporting); Project administration (lead); Resources (lead); Supervision (equal); Writing - review \& editing (equal).
Jens Petersen: Conceptualization (equal); Formal analysis (supporting); Methodology (lead); Software (supporting); Supervision (equal); Visualization (supporting); Writing - original draft (equal), Writing - review \& editing (equal).

\nocite{*}
\bibliography{hortensia}

\end{document}